# PHASE-FIELD MODELING OF CONTACT MELTING IN BINARY ALLOYS

Kartuzov V.V. [1], Bystrenko O.V. [1,2]

[1] Frantsevich Institute for material science problems, Kiev, Ukraine
[2] Bogolyubov Institute for theoretical physics, Kiev, Ukraine

**Abstract**. Computer simulations of the phenomenon of contact melting in binary alloys with chemical miscibility gap are performed on the basis of the phase field theory. Kinetics of the process is examined within the isothermal approximation, as a function of initial state. As evidenced by simulations, the simplest phase field model is capable of reproducing the basic properties of the phenomenon. The numerical results obtained suggest the diffusive nature of contact melting.

1. **Introduction.**

The goal of this work is a study of the phenomenon of contact melting (CM) characteristic of binary systems with chemical miscibility gap in solid state. Its essence is that the two solid components of a binary alloy, being separately in equilibrium at a temperature below both melting points, but above the minimum melting temperature, start melting when brought into contact. In spite of the fact that CM is rather common in multicomponent systems and has important industrial implications [1], its theoretical explanation is till now mainly limited to general thermodynamical considerations ignoring the kinetics of the phenomenon. To investigate CM in more detail, we employ computer simulations based on the phase-field theory (PFT), which received in recent years wide acceptance as a tool for description and numerical simulation of processes of structure- and phase formation in various materials [2-4]. PFT is based on the idea to describe the properties of complex systems in terms of continuous phase variables, which specify the phase state (liquid or solid) as well as any other properties of media. This makes possible to approximate the sharp interfaces between the phases by transition layers and avoid the complications associated with solving free boundary problems. To describe the microstructural evolution of a material, the dissipative dynamics equations for the phase fields, temperature and concentration are employed. It should be mentioned that the vast majority of works reporting PFT-based simulations (see, for instance, [2-7])are mainly dealing with the formation of various patterns in materials, while the above mentioned phenomenon of CM remains poorly examined.

In this work we use the simplest version of PFT for binary alloys, where the specific kind of phase diagram is provided by the chemical miscibility gap of solid components. In spite of limitations in the description of differences in solid phases and solid-solid interfaces [8], it still reproduces a number of basic properties of such alloys, and may provide a basis for important conclusions.

2. **Basic PFT equations.**

PFT is based upon the fundamental principles of irreversible thermodynamics. In particular, in order to study isothermal processes in binary systems, we can start with the free energy representation of a system in the form of the functional of phase variables and concentration

$$F = \int \left[ f(\varphi, c, T) - \frac{\varepsilon_C^2}{2}(\nabla c)^2 - \frac{\varepsilon_\Phi^2}{2}(\nabla \varphi)^2 \right] dV \qquad (1)$$

where $f(\varphi, c, T)$ is the free energy density; $T$, $c$, and $\varphi$ denote the temperature, concentration and the phase field, respectively. Then, by applying the fundamental requirement of thermodynamics that the free energy can only decrease in an isothermal process, one arrives at the PFT equations describing the dissipative microstructural dynamics in complex media. For the particular case of an isothermal binary system consisting of A and B components, they read [4]

$$\frac{\partial \varphi}{\partial t} = -M_\Phi \left[ \frac{\partial f}{\partial \varphi} - \varepsilon_\Phi^2 \nabla^2 \varphi \right] \qquad (2)$$

$$\frac{\partial c}{\partial t} = \nabla \cdot \left[ M_C c(1-c) \nabla \left( \frac{\partial f}{\partial c} - \varepsilon_C^2 \nabla^2 c \right) \right] \qquad (3)$$

Here the quantities $M_\Phi$ and $M_C$ are the kinetic coefficients specifying the relaxation rate of the system to the equilibrium state; $\varepsilon_\Phi$ and $\varepsilon_C$ are the model parameters, which determine the surface energy for interfaces and introduce into the model the dependence on the concentration and phase gradients; and the component concentrations $c_A$ and $c_B$ are related to the concentration $c$ as $c_A = 1-c$; $c_B = c$. The meaning of the phase variable is that it specifies the phase state, $\phi=0$ for solid, and $\phi=1$ for liquid, at the point $(x,t)$.

The specific features of the microstructural processes in the system are determined by the particular form of free energy density $f(\phi, c, T)$. In what follows, we use its classical expression for a non-ideal binary system [4]

$$f(\phi, c, T) = (1-c) f_A(\phi, T) + c f_B(\phi, T) + \frac{RT}{v_m} [c \ln c + (1-c) \ln(1-c)] \\ + c(1-c) [\Omega_S [1-p(\phi)] + \Omega_L p(\phi)] \qquad (4)$$

where $R$ is the gas constant, $v_m$ is the molar volume, $\Omega_S$ and $\Omega_L$ are the mixing energies for the solid and liquid state, respectively. The component free energies $f_A(\phi, T)$ and $f_B(\phi, T)$ have the form

$$f_{A,B}(\phi, T) = W_{A,B} g(\phi) + L_{A,B} \frac{T_M^{A,B} - T}{T_M^{A,B}} p(\phi) \qquad (5)$$

where $T_M^A$ and $T_M^B$ are the melting temperatures, $W_A$ and $W_B$ are the energetic barriers associated with the surface energy of liquid-solid interfaces for the components A and B; $L_A$ and $L_B$ are the component latent heats; the functions $g(\phi) = \phi^2(1-\phi)^2$ and $p(\phi) = \phi^3(10 - 15\phi + 6\phi^2)$ are the barrier and interpolating functions constructed in such a manner as to provide the description of the liquid-solid interfaces of a finite width. The kinetic coefficients $M_\Phi$ and $M_C$ are determined for a binary system as

$$M_\Phi = (1-c) M_A + c M_B \qquad (6)$$

$$M_C = \frac{D_S + p(\phi)(D_L - D_S)}{RT/v_m} \qquad (7)$$

with $D_S$ and $D_L$ being the solid and liquid diffusivities, respectively.

### 3. PFT modeling of CM. Numerical Results.

To examine the process of CM in a binary system within the isothermal approximation, the system of Eqs.(2-3) was solved numerically, in 1D and 2D geometry, for the following set of model parameters: domain of solution was $X_{max} = 2.0 \cdot 10^{-4}$ cm; time interval for solution of non-steady problem $t=0\ldots 5$ sec; difference in the mixing energies for solid and liquid phases $\Omega = 3500$ J/cm$^3$; melting temperatures $T_M^A = 3500$ and $T_M^B = 3000$ °K; latent heats $L_A = 4500$ and $L_B = 1000$ J/cm$^3$; surface energies for liquid/solid interfaces $\sigma_A = \sigma_B = 3 \cdot 10^{-5}$ J/cm$^2$; kinetic parameters for interfaces $\mu_A = 0.5$ cm/(°K sec); $\mu_B = 0.1$ cm/(°K sec); diffusivities $D_S = 10^{-8}$ cm$^2$/sec and $D_L = 10^{-5}$ cm$^2$/sec.

We give below the results of simulations in the dimensionless form, by using the basic units defined as follows. As the length unit $l_0$ we take the size of solution domain, for the time unit we use the typical diffusive time $t_0 = l_0^2 / D_S$; the quantities with the dimensionality of specific energy like $f(\phi, c, T)$, $\Omega$, $W_{A,B}$ and $L_{A,B}$ are measured in the units of $f_0 = RT_M^A / v_m$, and the temperature unit is $T_M^A$. Further dimensionless quantities (marked with overline) entering the basic PFT equations are defined as $\overline{D}_L \equiv D_L / D_S$; $\overline{M}_{A,B} \equiv M_{A,B} f_0 t_0$; $\overline{\varepsilon}_{\Phi,C} \equiv \varepsilon_{\Phi,C} / (l_0^2 f_0)$. Accordingly, in dimensionless form the relevant parameters used in PFT-simulations were: $\overline{T}_M^B = 0.86$ ($\overline{T}_M^A \equiv 1$); $\overline{L}_A = 1.15$, $\overline{L}_B = 0.25$; $\overline{\Omega} = 0.89$; $\overline{D}_L = 10^3$ ($\overline{D}_S \equiv 1$); $\overline{W}_A = \overline{W}_B = 2.3 \cdot 10^{-3}$; $\overline{M}_A = 10^8$; $\overline{M}_B = 7.8 \cdot 10^7$; $\overline{\varepsilon}_\Phi^2 = 1.15 \cdot 10^{-5}$; $\overline{\varepsilon}_C^2 = 1.15 \cdot 10^{-4}$.

Let us say a few words about the choice of parameters. Most of them are within the range typical for real materials and similar PFT simulations. The estimation of the concentration gradient coefficient $\varepsilon_c$ from the experimental data seems to be impractical. We assumed its square to be by an order greater than that of phase gradient coefficient $\varepsilon_\phi$, because the solid-solid surface energy is typically greater than that of solid-liquid interface, and the above coefficients are responsible for these interfacial energies.

In evaluating the parameters needed for simulations, we used the relations given in Ref.[4] between the barrier $W_{A,B}$ and surface $\sigma_{A,B}$ energies, respectively, as well as those for the kinetic parameters $\mu_{A,B}$ and $M_{A,B}$.

To examine the processes of CM, the relevant phase diagram with melting lines is needed. Conventionally, these can be obtained from the requirement that the chemical potential must be constant at solid-liquid interface [4]. Here we compute the phase diagram numerically, in a straightforward manner. For this purpose, PFT equations (2-3) were repeatedly solved with different initial conditions corresponding to high density set of points on the (*T,c*) plane in the vicinity of melting lines, and the final equilibrium solutions were used to identify the associated (solid, liquid, or mixed) phase state. The results for the above parameters are given in Fig.1.

In numerical simulations of CM Eqs.(2-3) were solved for the given constant temperature $\bar{T}=0.75$ above the minimum melting temperature $\bar{T}_E=0.61$ on the phase diagram, and below the melting temperatures $\bar{T}_M^B=0.86$ and $\bar{T}_M^A \equiv 1$ of both components.

The initial concentration distribution has been set in such a way as to describe two pure components (c=0 and c=1) in solid state separated by some transition region (of the width $\approx 0.05$), where c varies from 0 to 1 (Fig.2, right). Notice, that at the temperature $\bar{T}=0.75$ the points c=0 or c=1 specify the thermodynamically stable solid phases of pure components.

It should be pointed out, that in the absence of fluctuations the phenomenon of CM is not observed due to the fact that the PFT equations by themselves allow an infinite existence of steady solutions describing the overheated solid state. The usual way to treat this problem is to add the Gaussian noise terms responsible for fluctuations to PFT equations [9-11]. Unfortunately, the accurate description of dynamical fluctuations in PFT-formalism seems to be yet not solved theoretical problem (see, for instance, the comments on closely related issues in Refs. [12-13] and the references therein). For this reason, to trigger the process of CM, we introduced small additional noise with 0.01 mean-square deviation to the initial phase distribution ($\phi=0$) only. It is to mention that the amplitude of the initial noise considerably affects the initial starting time point, but not the kinetics (i.e., rate, typical behavior, etc.) of the CM process itself.

The boundary conditions used in simulations were $\frac{\partial c}{\partial x}=0\big|_{x=0,X\max}$ ; $\frac{\partial \phi}{\partial x}=0\big|_{x=0,X\max}$.

A number of computer runs with different initial conditions has been performed: a) 1D simulations for the average concentration $\bar{\bar{c}}=0.5$ with the corresponding point **Q** in the liquid area on the phase diagram (Fig.3); b) 1D simulations for the average concentration $\bar{\bar{c}}=0.2$ with excess of one of the components and with the corresponding point **P** in the solid area on the phase diagram (Fig.4); c) 1D simulations for the average concentration $\bar{\bar{c}}=0.36$ with the corresponding point **R** located in the area of liquid-solid coexistence (Fig.5); d) 2D simulations for the average concentration $\bar{\bar{c}}=0.5$ (Fig.6).

As is evidenced by the results obtained, the simulations reproduce the general features of CM; the latter manifests itself by the formation of quickly growing liquid area with $\phi=1$ (Figs.3-6). The short initial stage in all the cases is quite similar, while the final state of a system depends on the initial conditions used. In the case a), as one would expect, the complete melting of both components in the end is observed. In case b), the phenomenon of CM is observed as well, however, the diffusion processes give rise to equalization of the overall concentration resulting in the re-crystallization of the alloy into the final solid state with the concentration corresponding to the point **P** ($\bar{\bar{c}}=0.2$). In the case c), the final state turns out to be a mixture of solid and liquid of intermediate concentrations (c=0.22 and 0.44) indicating that the point **R** belongs to the area of solid-liquid coexistence. In Fig.1, this area, bounded by the solidus and liquidus lines, is filled in grey.

Results of 2D simulations correlate well with 1D case; however, as is seen from the figure, due to the effects of random noise, the process of CM starts non-simultaneously at different boundary points, which results in the irregularity of the liquid layer between the components. The distinguishing common feature of the kinetics of CM in all cases is a very short time of formation of liquid layer at initial stage, by orders less than the typical diffusive time $t_0 = l_0^2 / D_S$ used as the unit in simulations.

4. Conclusions

To conclude, the computer simulations performed have demonstrated the capability of the simplest PFT version with chemical miscibility gap to reproduce the basic properties of CM in binary alloys. After the very short initial stage associated with the formation of liquid layer between the solid components, the melting proceeds to a steady state, which depends on the parameters, phase diagram and initial state of the system. The results obtained suggest the diffusive nature of contact melting, since the latter is observed within the isothermal approximation.

Acknowledgements

The authors acknowledge the support of EOARD, Project No 118003 (STCU Project P-510).

References

1. Zalkin V M 1987 *Nature of Eutectic Alloys and Effect of Contact Melting* (Moscow: Metallurgiya).
2. Singer-Loginova I, Singer H M 2008 Rep. Prog. Phys. **71** 106501.
3. Chen L Q 2002 Ann. Rev. Mater. Res. **32** 113–140.
4. Boettinger W J, Warren J A, Beckermann C and Karma A 2002 Annu. Rev. Mater. Res. **32** 163.
5. Apel M, Boettger B, Diepers H J, Steinbach I 2002 J. of Crystal Growth **237–239** 154–158.
6. Nestler B, Wheeler A A 2002 Computer Physics Communications **147** 230–233.
7. Lewis D, Pusztai T, Gránásy L, Warren J and Boettinger W 2004 JOM **April** 34-39.
8. Wheeler A A, Mc Fadden G B and Boettinger W J 1996 Proc. R. Soc. Lond. **452** 495-525 (no. 1946).
9. Elder K R, Provatas N, Berry J, Stefanovic P and Grant M 2007 Phys. Rev. **B75** 064107.
10. Drolet F, Elder K R, Grant M and Kosterlitz J M 2000 Phys.Rev. **E61** 6705.
11. Karma A and Rappel W J 1999 Phys. Rev. **E60** 3614.
12. Van Teeffelen S, Backofen R, Voigt A and Löwen H 2009 Phys. Rev. **E79** 051404.
13. Marconi U M B and Tarazona P 1999 J. Chem. Phys. **110** 8032.

**Figures**

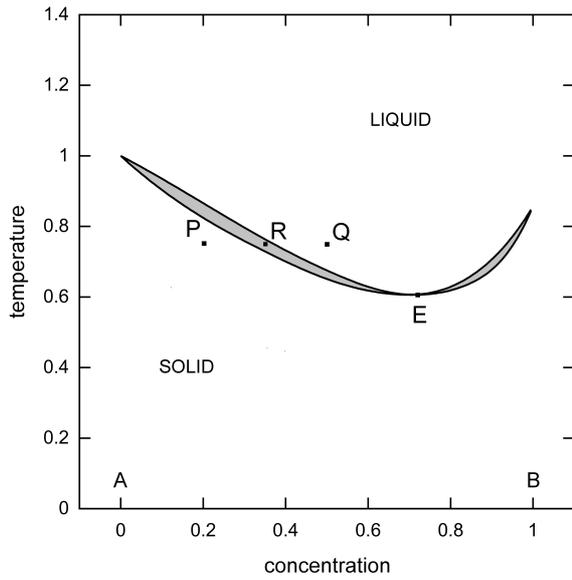

Fig.1. Phase diagram for a binary system obtained from PFT equations for $\bar{T}_M^B =0.86$ ( $\bar{T}_M^A \equiv 1$ ); $\bar{L}_A =1.15$ and $\bar{L}_B =0.25$; $\bar{\Omega} =0.89$. Liquid-solid coexistence area is filled in grey.

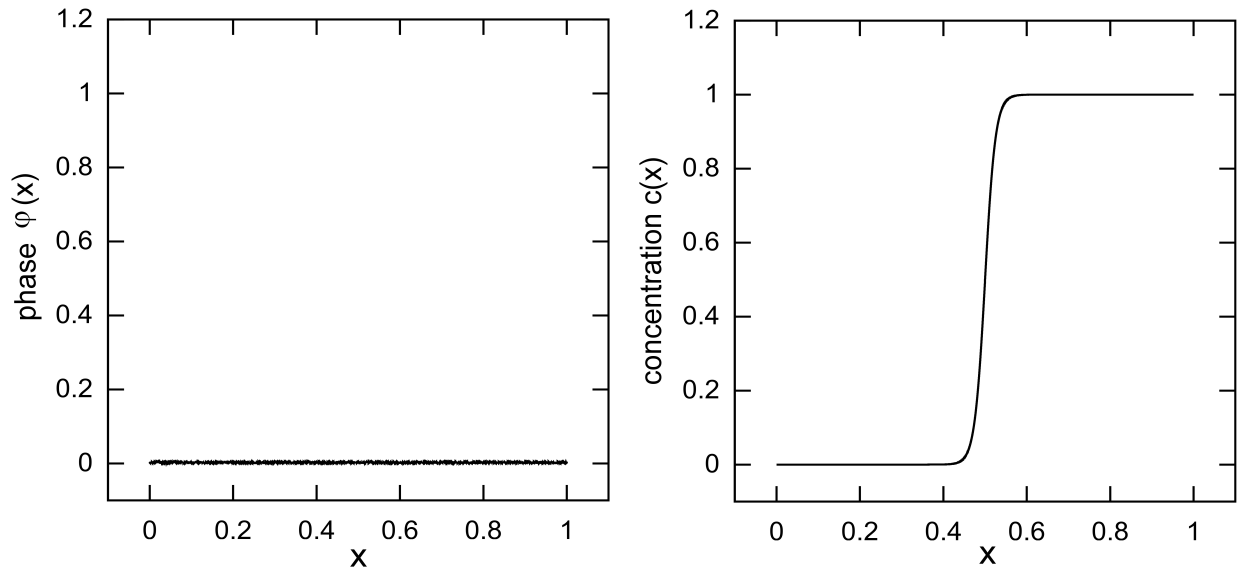

Fig.2. Typical initial distributions for the phase (left) and concentration (right). The initial distribution of phase variable corresponds to the solid state $\varphi=0$ with 0.01-amplitude noise.

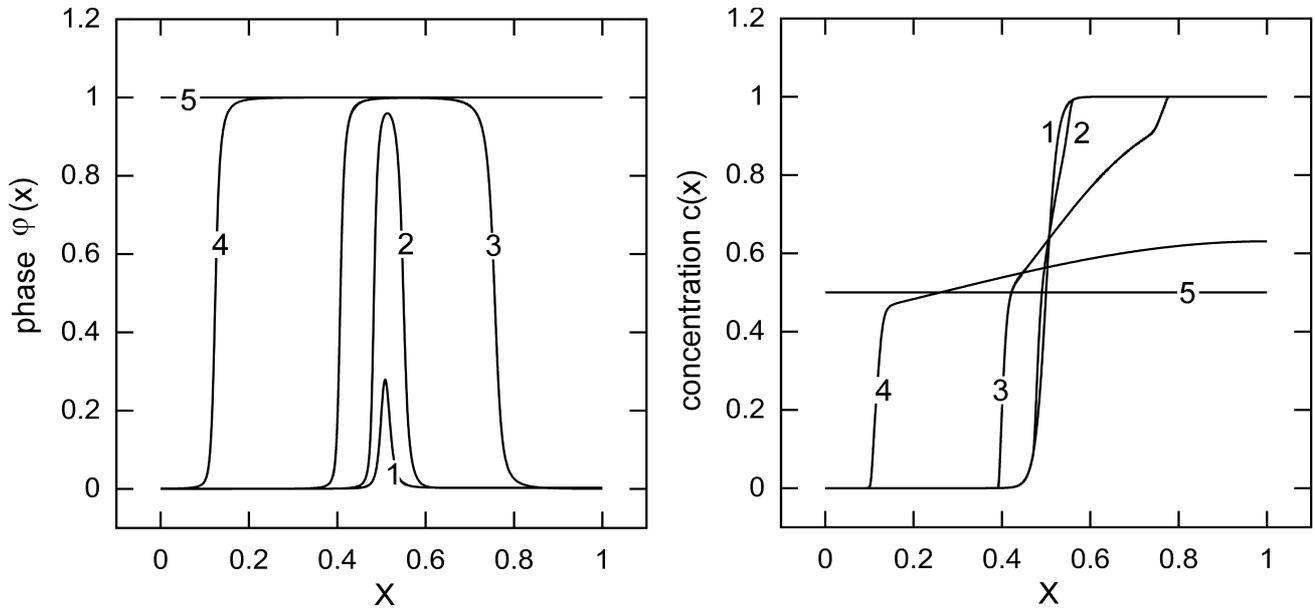

Fig.3. Kinetics of CM process for the average concentration $\bar{\bar{c}}=0.5$ (point Q) ; the phase (left) and concentration (right) curves are given for the time points $t=1.7 \cdot 10^{-6}(1); 2.3 \cdot 10^{-6}(2); 2.0 \cdot 10^{-5}(3); 3.0 \cdot 10^{-4}(4); 1.0(5)$. In the end, the complete melting ( $\phi=1$ ) is observed in the system.

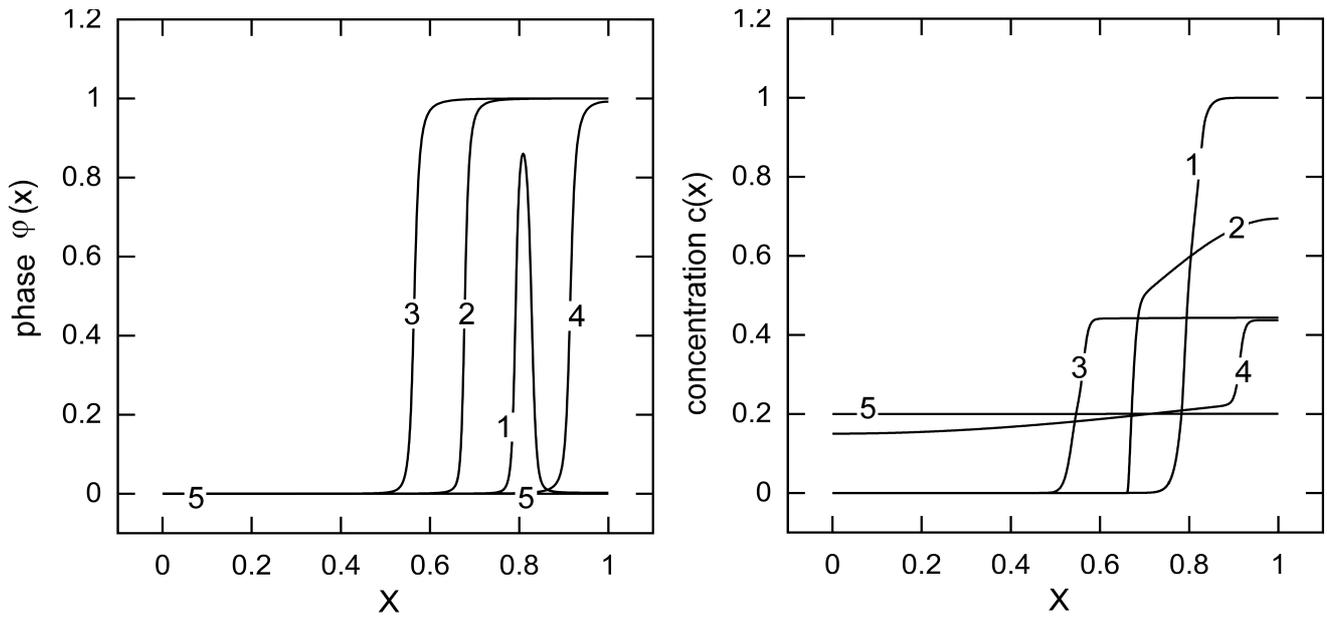

Fig.4. Phase (left) and concentration (right) evolution in the process of CM for the average concentration $\bar{\bar{c}}=0.2$ (point P) for the time points $t=2.0 \cdot 10^{-6}(1); 3.3 \cdot 10^{-5}(2); 5.0 \cdot 10^{-4}(3); 0.4(4); 1.0(5)$. In the end, the complete re-solidification ( $\phi=0$ ) of the melt formed in the process of CM is observed.

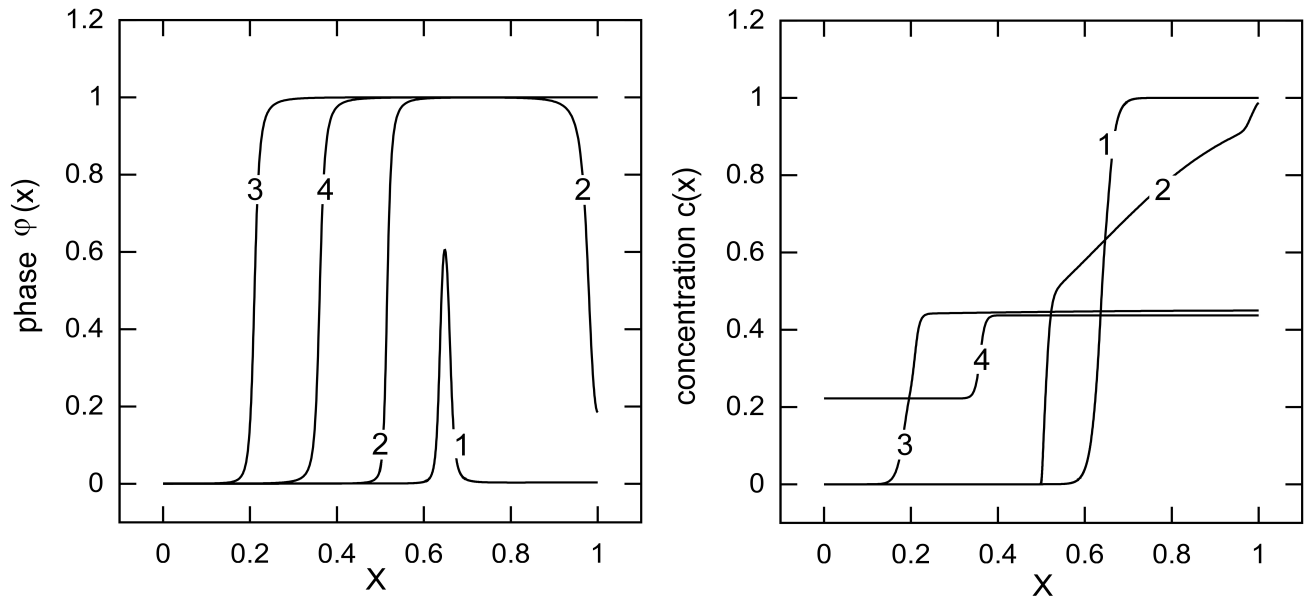

Fig.5. Phase (left) and concentration (right) evolution in the process of CM for the average concentration $\bar{\bar{c}}=0.36$ (point R) for the time points $t=1.7 \cdot 10^{-6}(1);\ 3.3 \cdot 10^{-5}(2);\ 1.0 \cdot 10^{-3}(3);\ 1.0(4)$. In the final state, the co-existence of solid ($\phi=0$ at $c=0.22$) and liquid ($\phi=1$ at $c=0.44$) is observed.

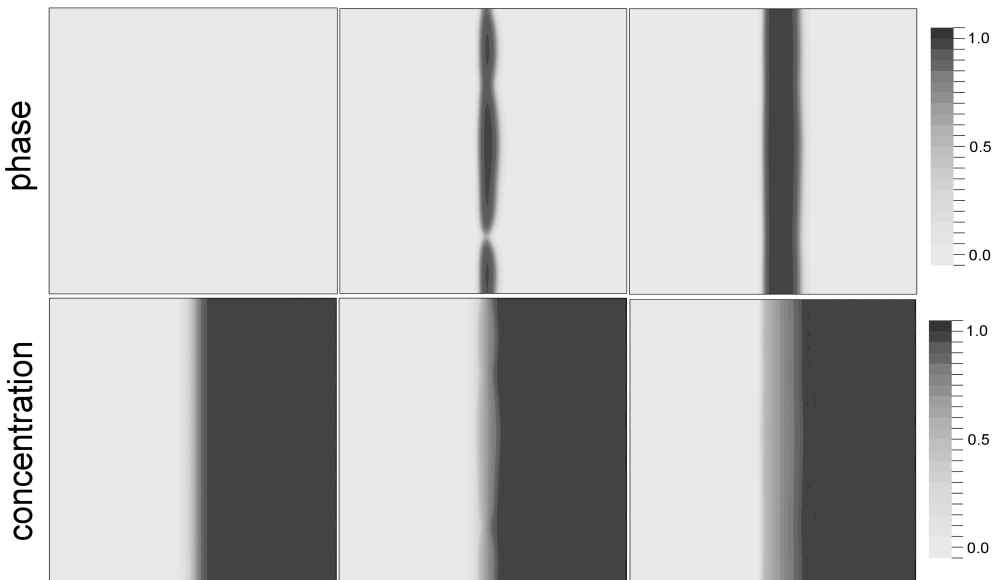

Fig.6. Results of 2D PFT simulation of CM for the average concentration $\bar{\bar{c}}=0.5$ (point Q). Evolution of phase (top) and concentration (bottom) distributions on the unit square for $t=0;\ 2.0 \cdot 10^{-6};\ 4.0 \cdot 10^{-6}$.